# JC-PIC: kinetic (PIC-MCC) simulation of low-temperature plasmas with a documented library of one hundred test cases

**Jean-Pierre Boeuf** — CNRS Emeritus Senior Research Scientist, LAPLACE, CNRS, Université de Toulouse, France

**Abstract.** Over the last twenty years, particle-in-cell simulations with Monte Carlo collisions (PIC-MCC) have given the low-temperature plasma community much of what it knows about electron heating in RF discharges, sheaths and secondary emission, striations, and the instabilities of magnetised plasmas. But these results remain in papers, and in codes that only their authors can run. JC-PIC is a free Windows program: a one-dimensional, three-velocity (1D3V) electrostatic PIC-MCC code with a parallel Fortran engine (OpenMP) and a complete graphical interface. It is built around a library of about one hundred ready-to-run cases, each documented, referenced and explained. About half of them reproduce a published result under the original conditions; the others compare the code with a classical analytical result. This overview presents the program, its companion book, and six of the cases from the physics point of view: Langmuir waves from Landau damping to electron trapping, the heating-mode transition in argon RF discharges, the electrical asymmetry effect, the Franck–Hertz experiment treated as a swarm, the striations of a positive column, and the E×B drift instability of Hall thrusters.

## 1. What JC-PIC is, and why

The Particle-In-Cell method represents a plasma by a large number of charged super-particles. They move in the electric field computed on a mesh from their own charge. With a Monte Carlo treatment of the collisions with the neutral gas (PIC-MCC), the method is kinetic: it is equivalent to solving the Boltzmann equations of electrons and ions together with Poisson's equation, and it is the reference against which fluid and global models are judged. An ordinary desktop computer can now follow millions of particles, so a one-dimensional kinetic simulation of a real discharge takes minutes to hours. One thing has not changed: a PIC-MCC code is a research tool, written and run by modellers.

JC-PIC was written to remove this barrier. It is a 1D3V electrostatic PIC-MCC code: one space coordinate, across a plasma column or between two electrodes, and the three velocity components of each particle, so that magnetic fields and the angular distribution of collisions are treated correctly. The numerical scheme is standard: explicit leapfrog, Boris pusher, null-collision Monte Carlo, tabulated cross sections in the LXCat format. What is unusual is the environment built around the engine. The user sets up a discharge in a dialog: gas, pressure, gap, voltage waveform or imposed current, magnetic field, walls and secondary emission. The simulation runs in the background on all the cores of the machine, and the user watches it evolve: density and potential profiles, electron energy probability functions in the bulk and at the walls, ion flux-energy distributions, position–time maps of any quantity over the RF period, phase space, currents and their spectra, and the decomposition of the electron power in the manner of Schulze et al. Every viewer exports its figure and its numbers. A paused run can be resumed, even on another machine. Nothing requires a line of code or a script.

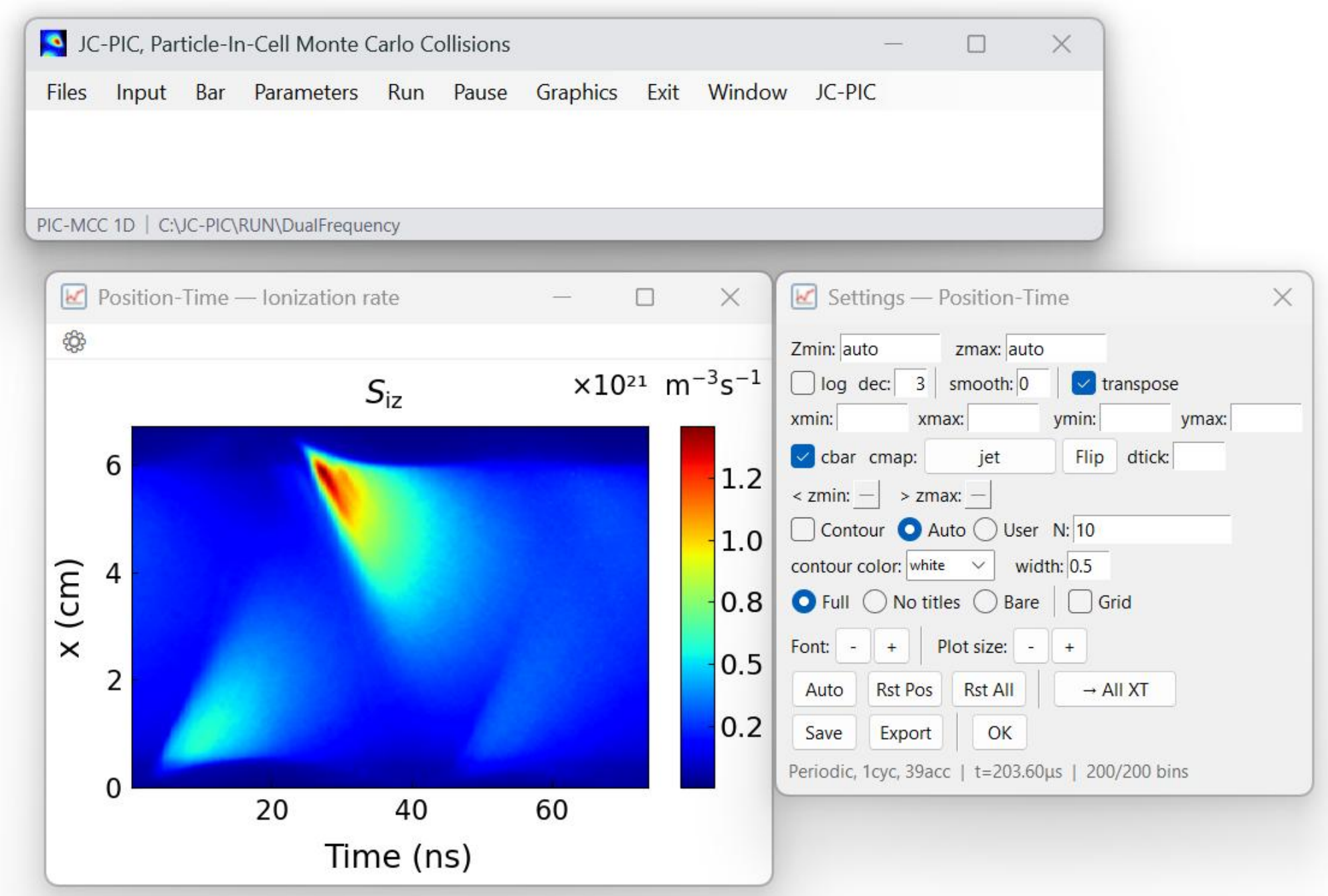

Figure 1. The JC-PIC main window and one of its viewers: the position–time map of the ionization rate over one RF period in a dual-frequency capacitive discharge (the electrical asymmetry effect, section 3.3). The viewers reload their data from disk at each refresh, so a run can be examined at any time while it computes.

The project has two aims, which support each other:

- The first aim is educational and exploratory: to put a real kinetic simulation in the hands of experimentalists, students and teachers, so that the key results of the modern PIC-MCC literature can be reproduced, examined and understood directly.
- The second aim is scientific: to look again at the physics of low-temperature discharges in the light of these simulations, and to gather in one place, and check one by one, the main conclusions of a large body of published work.

A third element should be stated, because it made the other two possible. The Fortran engine, the graphical interface, the diagnostics and the case descriptions were produced by a single author with the assistance of Anthropic's Claude, under the author's scientific direction: the physics specifications, the choice of the test cases, the validation against the literature and the interpretation of every result are the author's. Work of this size would ordinarily have taken years; it took a few months. That experience is itself part of what the project has to report.

The present manuscript was written by the author. Every reference in the list below was checked against the original paper, and every number quoted in what follows was measured in the runs distributed with the library.

## 2. The test-case library and its companion book

The heart of JC-PIC is not the engine but the library that comes with it: about one hundred ready-to-run cases, organized by topic and browsed from the program (Figure 2). Each case is a folder with its input file, its stored results, and a description written for a reader, not for a machine: the physics in question, the simulation conditions, the results — compared with the original paper when the case reproduces one, or with the analytical result when it tests one — and the references. Loading a case copies it to a working folder. Its plots open immediately from the stored results. Pressing Run continues the simulation from its converged state, or starts it again from the beginning. The user can then change one parameter — the pressure, the voltage, the magnetic field — and see what changes, in a few minutes.

The topics follow the physics: **basic plasma physics** (sheath and plasma potential, the sheath in an oblique magnetic field, ambipolar diffusion, Langmuir and ion-acoustic waves, plasma expansion, the link with the global model); **swarm physics** (transport coefficients, pulsed and steady-state Townsend experiments, the Franck–Hertz experiment); **electron emission** (Child–Langmuir and thermionic diodes, virtual-cathode oscillations, thermionic discharges and their anode-glow mode); **DC and transient glow discharges** (helium glow, the Carlsson benchmark, plasma-immersion ion implantation); **capacitive RF discharges** (the Turner and eduPIC benchmarks, Godyak's pressure effect, dual-frequency electrical asymmetry, Schulze's power absorption, frequency effects, magnetised RF); **positive columns** (striations, local and non-local regimes, the Hall effect); **magnetised plasmas** (the Hall-thruster E×B instability); **kinetic instabilities** (two-stream, Buneman); and an **appendix** of definitions and interactive modules.

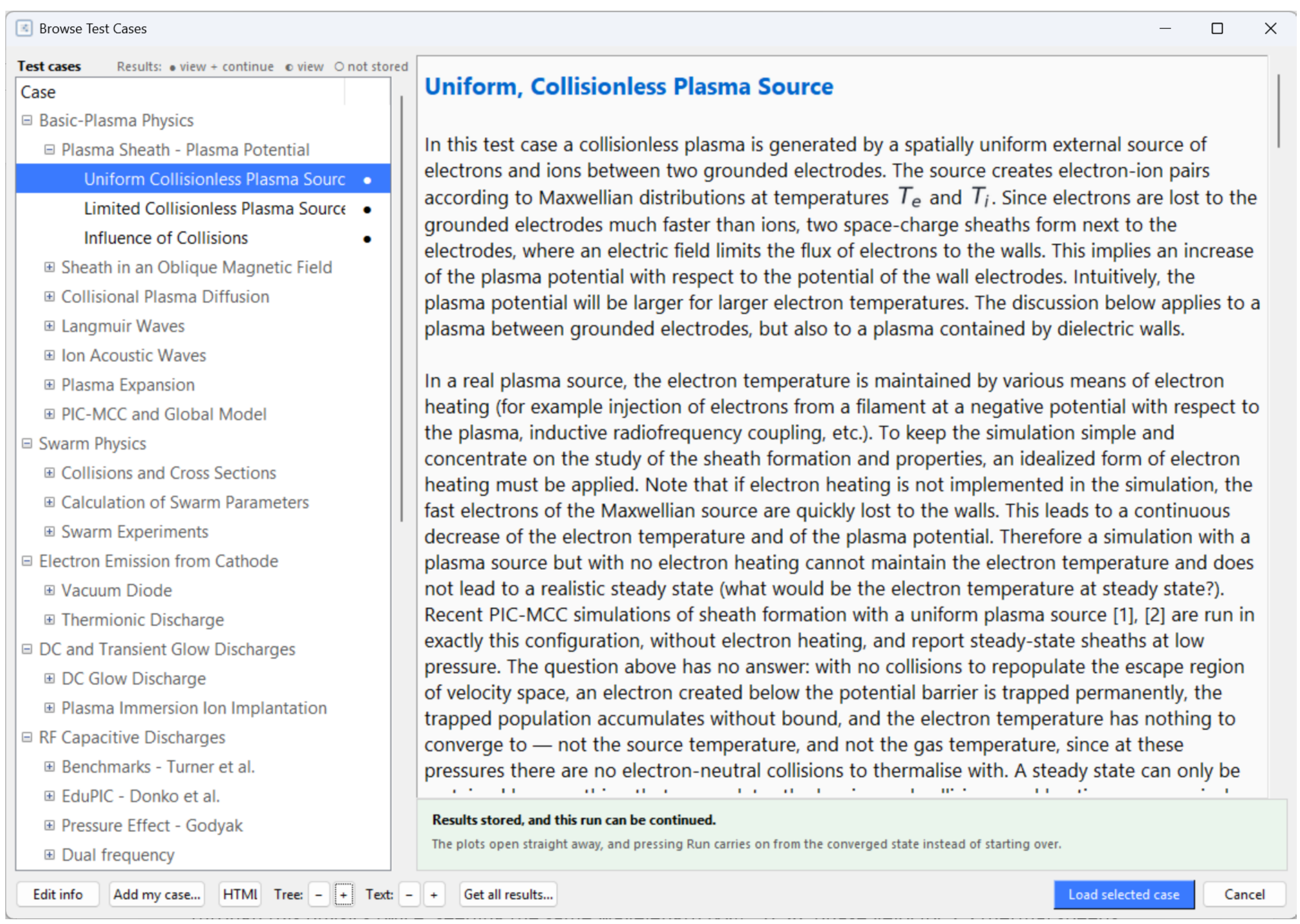

Figure 2. The case browser: the library tree on the left (with the state of the stored results of each case), the description of the selected case on the right. A user's own cases are added to the same tree, and can be sent to the author for inclusion in the library under the contributor's name.

The companion book, ***Physics of Low Temperature Plasmas via Particle Simulation*** (CC BY 4.0, doi:10.5281/zenodo.22258142), collects all the case descriptions with their figures and animations, and a single reference list of about 150 papers. It is written so that a reader can follow the physics without the software, and so that a user of the software can read the chapter, load the case, run it and see it. Six of its cases are presented below, from the physics point of view rather than the numerical one. The numbers quoted are those measured in the runs shipped with the library.

# 3. Six cases, seen from the physics

## 3.1 Langmuir waves: from Landau damping to electron trapping

The simplest motion of a plasma is an oscillation. Displace a slab of electrons on a fixed ion background: the charge separation pulls them back, they overshoot, and the plasma oscillates at the electron plasma frequency. With a finite electron temperature, the pressure couples neighbouring slabs and the oscillation becomes a wave, the Langmuir wave of Bohm and Gross [1]. Landau showed in 1946 [2] that such a wave decays without any collision: electrons slightly slower than the wave take energy from it, electrons slightly faster give energy back, and since a Maxwellian contains more slow than fast particles, the wave loses energy. The two cases of the library apply this physics twice to the same plasma ($2\times10^{14}$ m$^{-3}$, 3 eV, a 6 cm periodic box, no gas, immobile ions) and the same wavelength ($k\lambda_D$ = 0.38, phase velocity 3.3 thermal speeds, in the tail of the Maxwellian), with a density perturbation of 2 % in the first case and 30 % in the second.

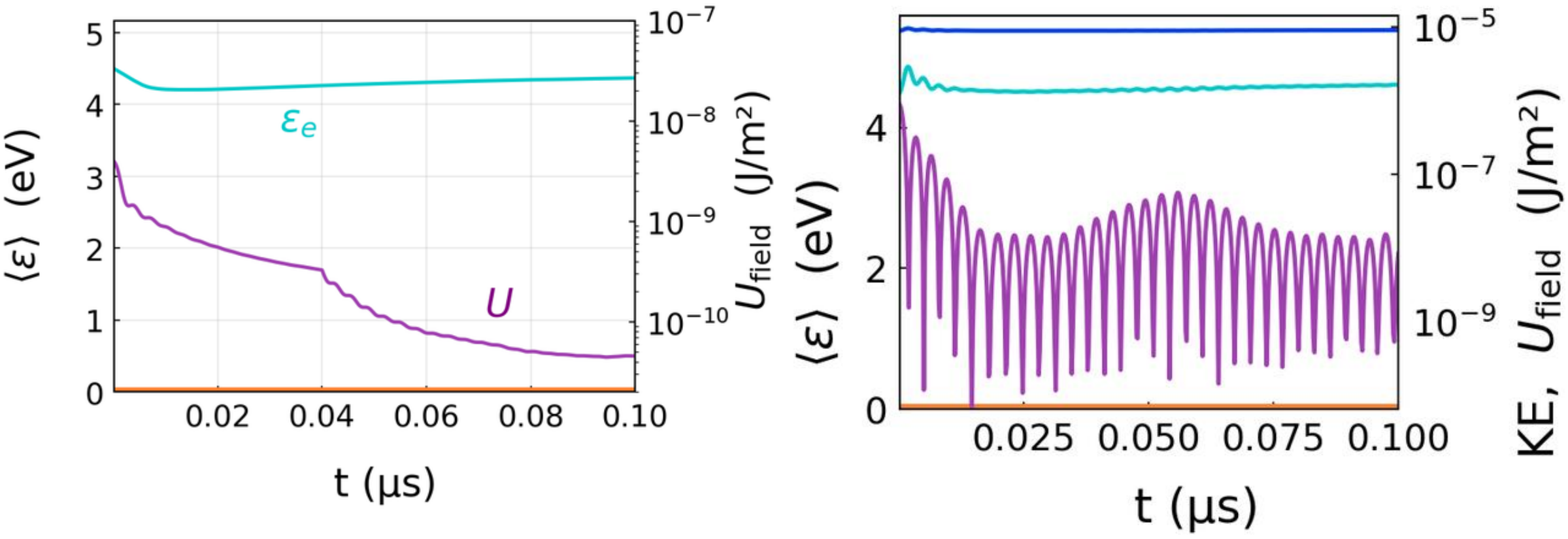

Figure 3. The electrostatic field energy (purple, log scale) against time, for the same plasma and the same wave, seeded at 2 % (left) and at 30 % (right). Left: a straight line, i.e. exponential Landau damping; the measured rate is 0.053 ω_pe, against 0.081 for the textbook asymptotic formula, and the frequency is 5 % above the fluid Bohm–Gross value; both are exactly the roots of the kinetic dispersion relation. Right: the energy falls by two decades in about 20 ns, rises again around 50–60 ns and falls again. This is the amplitude oscillation of O'Neil [3]: the wave and its trapped electrons exchange energy. The cyan curve is the electron mean energy: flat on the left; on the right it goes from 4.8 to 4.55 eV as the ordered oscillation of the electrons becomes disordered kinetic energy, at constant total energy (blue).

At 2 %, everything Landau wrote applies: the field energy decays as a straight line on a log plot, at the exact kinetic rate (the simulation is accurate enough to distinguish it from the textbook approximation), and the phase space remains a smooth Maxwellian band. At 30 %, the potential wells of the wave are 2 $T_e$ deep and the resonant electrons cannot pass through them. They bounce inside the wells, at a frequency ten times the Landau rate, and within one bounce period they roll the phase space into the vortices ("cat's eyes") of a Bernstein–Greene–Kruskal mode [4]: one per wavelength and per direction of propagation, since the seed splits into a standing wave (Figure 4). Landau damping stops, the wave survives four linear damping times, and the electron distribution develops a plateau between 15 and 45 eV with a sharp cut-off near 55 eV, the top of the trapping separatrix. A factor of fifteen on one number of the input file takes the same plasma from the linear regime, which a fluid model can describe, to the fully kinetic regime, which no fluid model can describe.

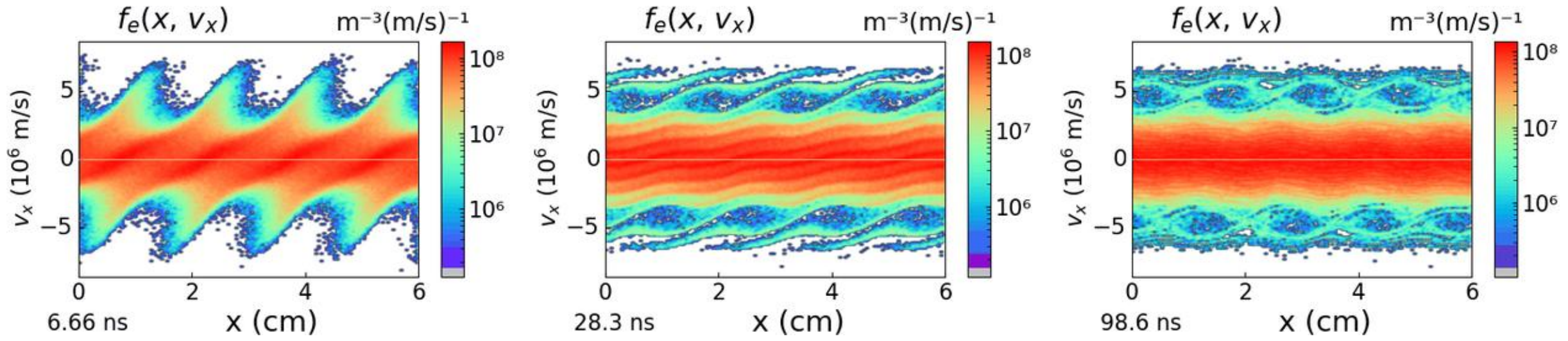

Figure 4. The 30 % case: electron phase space f(x, $v_x$) at three times: during the formation of the vortices (7 ns, half a bounce period); in the established BGK regime (30 ns, eight cat's eyes at ± the phase velocity, with the empty separatrix between trapped and passing electrons); and at the end of the run (100 ns, the vortices persist but their interiors have phase-mixed into spiral filaments).

## 3.2 Godyak's heating-mode transition: when an argon discharge gets colder at lower pressure

In 1990, Godyak and Piejak [5] measured the electron energy distribution in a symmetric argon RF discharge while lowering the pressure at fixed current. They found something a fluid model cannot produce: below about 50 mTorr, the effective electron temperature drops to a fraction of an electronvolt, and the distribution has two temperatures — a large cold group at a few tenths of an eV, and a hot, dilute tail. Helium, measured in the same cell, shows nothing of the kind. The reason is the Ramsauer minimum of the argon cross section near 0.3 eV. The slow electrons created by ionization in the centre have a collision frequency so low that they are trapped in the ambipolar potential well; they oscillate almost without collisions in the RF field and are not heated. The fast electrons bounce between the sheaths, are heated there stochastically, and produce the ionization. When the pressure is raised, this arrangement breaks down suddenly, because the feedback is positive.

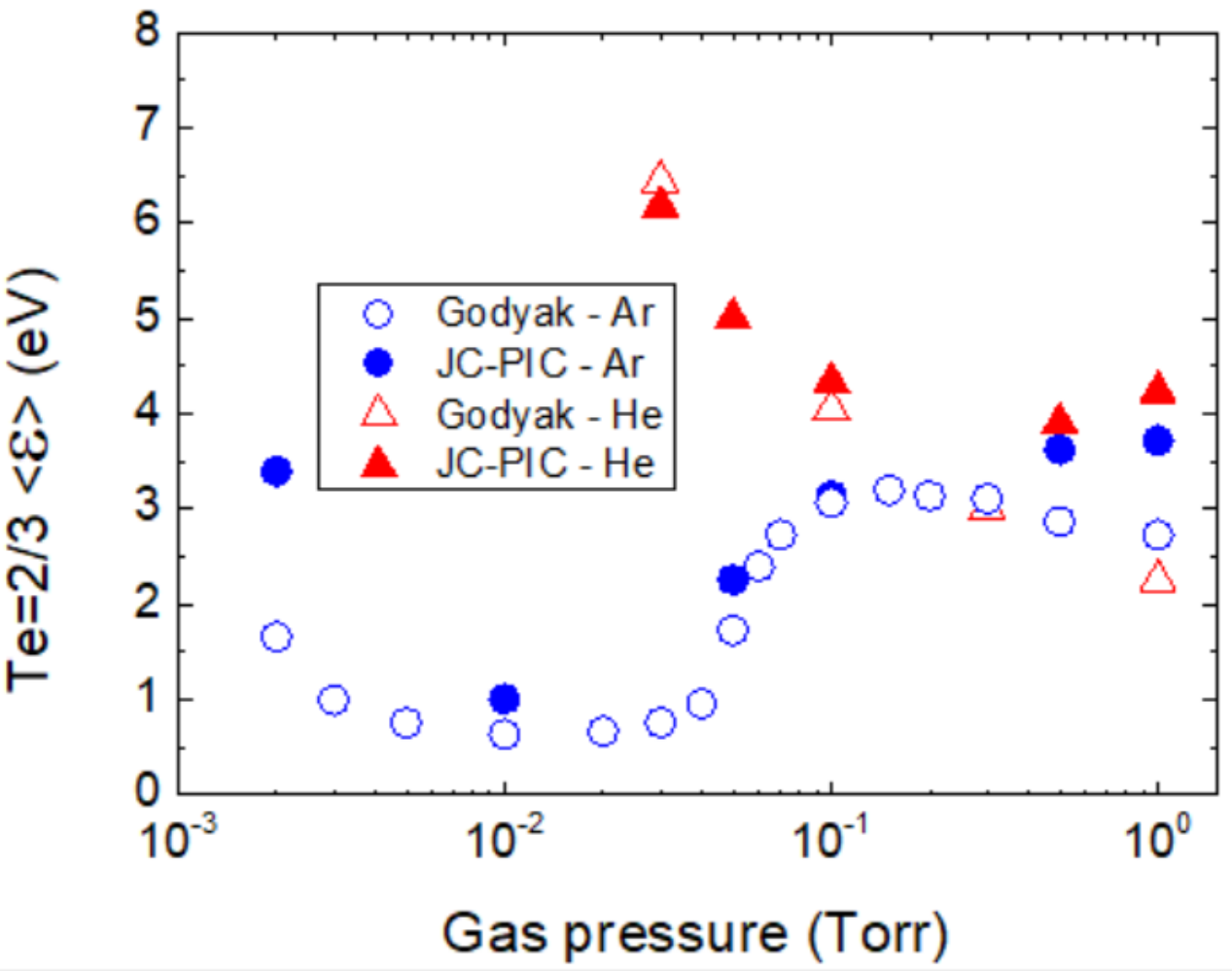

Figure 5. Effective electron temperature (two thirds of the mean energy at the midplane) against pressure, at a fixed current density of 10 A/m², 13.56 MHz, 6.7 cm gap. Open symbols: the measurements of Godyak et al. Filled symbols: JC-PIC, current-driven like the experiment. The non-monotonic behaviour of argon and the ordinary monotonic behaviour of helium are both reproduced, with the minimum at 10 mTorr, where it is measured.

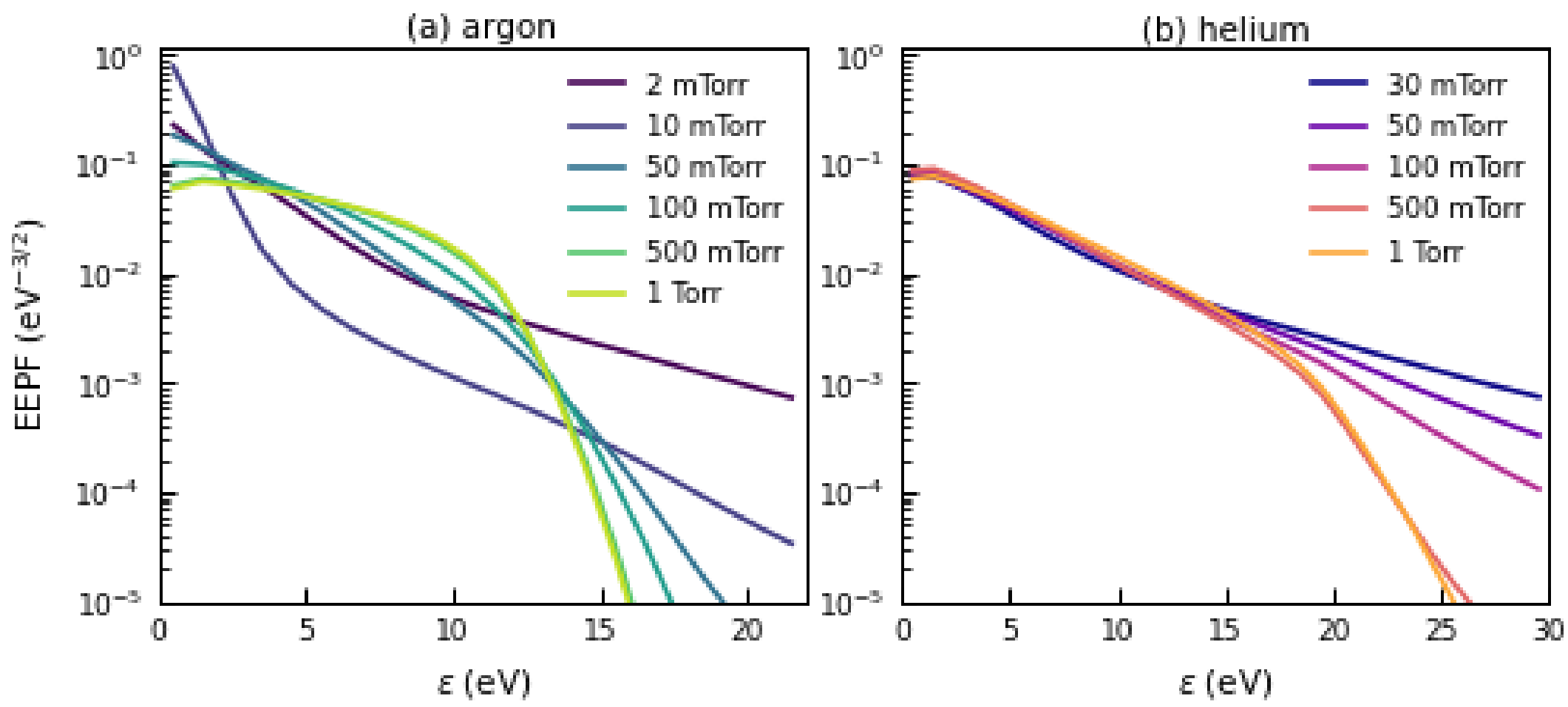

Figure 6. The electron energy probability functions at the midplane for the whole series, (a) argon, (b) helium. In argon the function is concave at low pressure — a steep cold part and a long hot tail, two temperatures — and convex above the transition. In helium it is monotonic at every pressure.

The library runs these discharges as Godyak did, with the current imposed rather than the voltage; JC-PIC does this exactly with its current-source drive. It reproduces the whole shape of the curve without any adjustment: the fall to a minimum, the position of the minimum, the steep rise between 40 and 100 mTorr, and the plateau. Where it departs from the measurement, the case says so. Below 50 mTorr the simulation is too hot by a factor 1.5–2, for a reason a user can test: the trapped cold group is destroyed by anything that scatters it, including numerical field noise (the 10 mTorr case comes with its convergence study). Above 500 mTorr it is too hot by 30 % in both gases; a missing loss channel (no excited-state population) is the most likely cause. A case that shows the limits of a model teaches more than one that hides them.

### 3.3 Two frequencies and one phase: the electrical asymmetry effect

In a capacitive discharge, the ion flux and the ion energy at the electrodes are linked: raising the voltage to increase the flux also increases the energy. Heil and Czarnetzki showed in 2008 [6] that a voltage waveform made of a fundamental and its second harmonic, $V(t) = V_0[\cos(2\pi ft + \theta) + \cos(4\pi ft)]$, makes the two sheaths different even in a perfectly symmetric reactor. The DC self-bias, and therefore the ion energy, can then be adjusted continuously with the phase θ alone, at constant flux. Donkó, Schulze, Heil and Czarnetzki confirmed the effect by PIC simulation in 2009 [7]. The library reproduces their phase scan in argon at 20 mTorr, 6.7 cm gap and 315 V per harmonic.

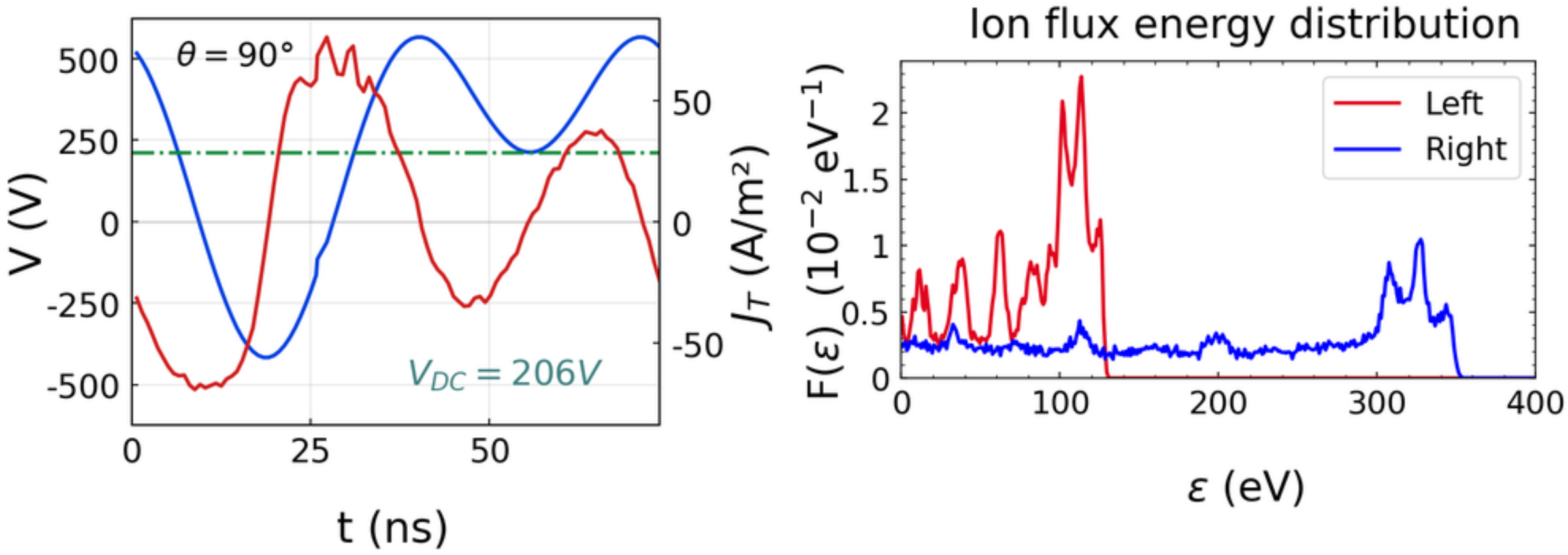

Figure 7. The θ = 90° case. Left: potential of the powered electrode and total current density over one 13.56 MHz period; the dash-dotted line is the DC self-bias, +206 V, against +215 V in Donkó et al. (and −212.5 V against −213 V at θ = 0°). Right: the ion flux-energy distributions at the two electrodes — 130 eV at most on the powered side, 353 eV on the grounded side, a factor 2.7 in energy for a difference of 9 % in flux. The fine double peaks are the bimodal RF structure of a collisional sheath; the flat continuum comes from charge exchange.

The agreement is quantitative: self-bias to 5 %, asymmetry parameter ε to 0.01. But the main point of the case is the shape of the two distributions: completely separated, and exchanged (mirror images) when θ goes from 0° to 90°, while the flux hardly changes. The position–time viewer shows where in the cycle the ionization takes place (Figure 1), and the current trace shows the plasma series resonance that Donkó et al. reported at the same phase of the period.

### 3.4 The Franck–Hertz experiment, seen as a swarm

The 1914 experiment of Franck and Hertz [8] is in every textbook: electrons are accelerated through a low-pressure vapour, the collected current drops at regular intervals of the voltage (4.9 V in mercury), and Bohr read this interval as the quantum of excitation of the atom. From the point of view of gaseous electronics, however, the Franck–Hertz tube is not a beam experiment. It is a Townsend discharge below the ionization threshold, in a strongly non-hydrodynamic regime: the electrons form a swarm, scattered elastically hundreds of times between two inelastic collisions, and the periodic structure that made the experiment famous is the spatial relaxation oscillation of their energy distribution. The library runs it in neon at 1 Torr, in a 40 cm gap, with a uniform field of 5 V/cm imposed rather than computed (space charge is negligible at these currents), and a 1 eV electron beam injected at the cathode. It looks at the structure along the gap instead of sweeping the voltage: in a uniform field the two views show the same variable, and the spatial view shows the whole distribution at once.

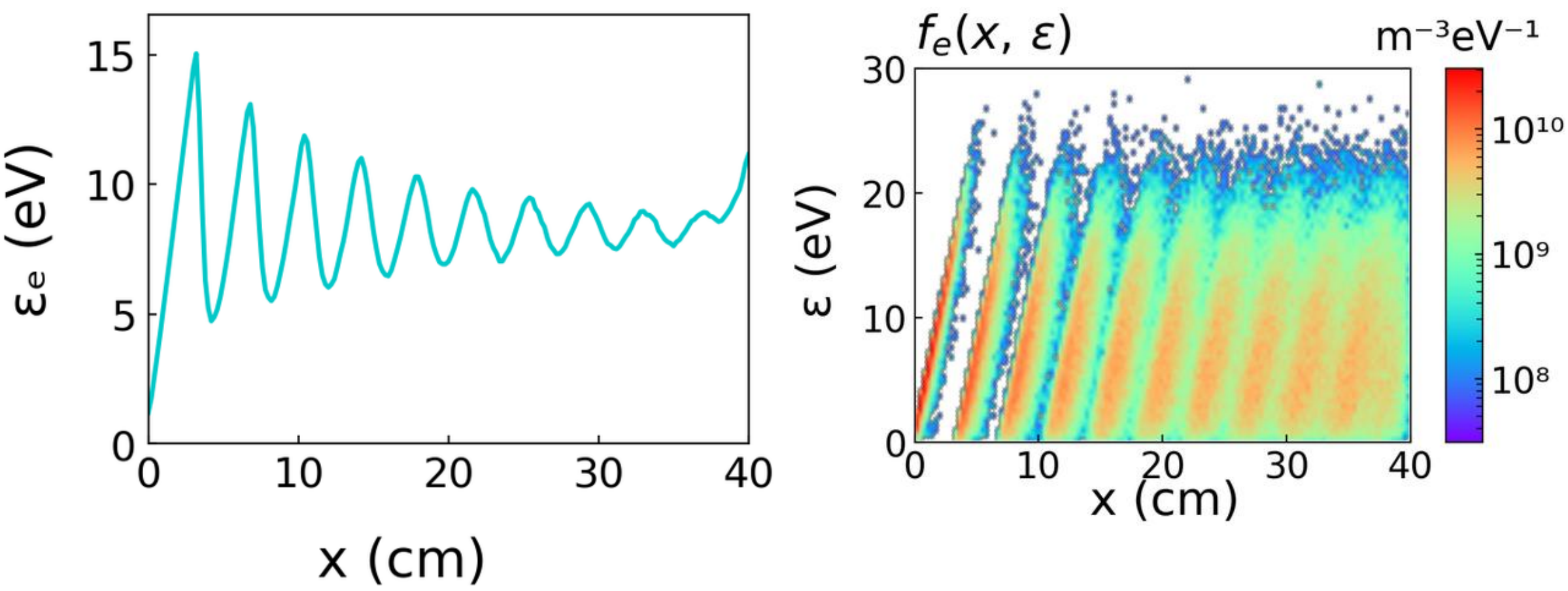

Figure 8. Neon, 1 Torr, 5 V/cm. Left: the mean electron energy along the gap. The swarm enters at 1 eV, rises to 15 eV at 3.2 cm, falls to 5 eV, and repeats with a period of 3.72 cm, i.e. 18.6 eV of potential drop, against the 18.7 V measured in real neon tubes and the 16.6 eV of the lowest threshold. Right: the electron distribution f(x, ε). Each straight branch is one generation of electrons accelerating freely at 5 eV/cm, cut near 20 eV by the inelastic collisions that bring them back to low energy. Successive generations broaden until, beyond 20 cm, the memory of the injection is lost.

Three results come out that the textbook picture does not contain. First, the period is not the excitation energy of the atom. It exceeds the lowest threshold by two volts, because the cross section rises slowly above threshold, because elastic scattering spreads the distribution, and because the 3p levels open at 18.4 eV; the simulation recovers the experimental 18.7 V from the cross sections alone. Second, the electrons are not a beam. The phase space is symmetric in $\pm v_x$ everywhere, an electron travels about one metre to advance by one period, and what is periodic in space is a local energy balance, not a trajectory; this is why the period scales as 1/E and does not depend on the pressure. Third, the oscillation is damped: the contrast in the excitation rate goes from three decades in the first luminous layer to a factor of two or three in the tenth. This is the phase mixing of successive generations, relaxing toward the ordinary hydrodynamic equilibrium of a swarm — 8.3 eV here, against 8.07 eV from the swarm module of JC-PIC at the same E/N. The luminous layers are also shifted 0.3 period downstream of the energy maxima, because light is a threshold quantity fed by the tail of the distribution, while the mean energy is a bulk average.

### 3.5 Striations: the same resonance, now self-organized

Positive columns at low pressure and low current rarely glow uniformly. The light comes in layers, standing or moving, and in rare gases the potential drop across one layer is a fixed fraction of the excitation threshold: one half, two thirds or one, the p, r and s waves of the experimental classification. The Franck–Hertz structure is this resonance, imposed on a swarm by the cathode. In a positive column the field is produced by the plasma itself, the structure acts back on the ionization that sustains it, and it becomes an instability. Fluid models reproduce the striations of the high-pressure, high-current part of the diagram, where Coulomb collisions and stepwise ionization make the ionization rate non-linear in the density. At low pressure and low current, none of this operates, the distribution is far from Maxwellian, and the mechanism is different: the column stratifies when the electron energy flux driven by the density gradient (the Dufour coefficient) is sufficiently negative [9]. This is a property of the shape of the distribution in the uniform field, before any perturbation. The library case is the simplest geometry where this can be studied: neon at 1 Torr, a 6 cm column closed on itself, an average field of 6 V/cm imposed along it, and wall losses represented by removing one electron–ion pair, at random, for each ionization.

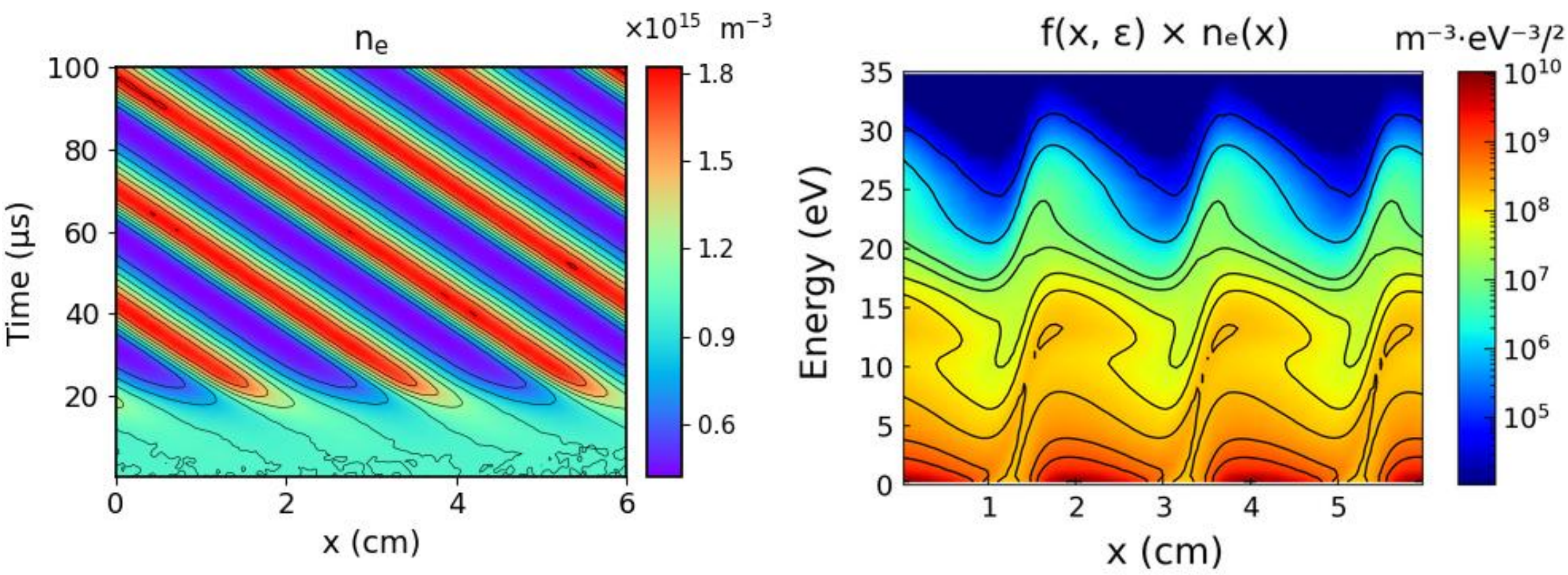

Figure 9. The DC neon column. Left: electron density in the position–time plane over the run. The column is uniform for ten microseconds; then the instability grows with an e-folding time of 5 μs and saturates into three striations that move toward the cathode at 0.75 km/s, a frequency of 37 kHz. Right: the electron energy probability function in the energy–position plane: slow electrons trapped in the ambipolar wells at the density maxima, and, between two maxima, a bump near 12 eV — the electrons released from a well that have fallen through the potential step of one striation.

In the saturated state, the column carries three striations of 2.0 cm, with the density modulated by a factor of four. The potential rises in steps of 12 V, close to two thirds of the neon threshold: an r wave. The mean electron energy varies by ±20 % across a striation, while the ionization rate varies by a factor of 240: the field modulates the tail of the distribution, not the bulk. The maxima of field, ionization and density do not coincide. The ionization peaks 2 mm on the cathode side of the density maximum, so new pairs appear ahead of the crest and the crest moves; this is what makes the wave travel, and sets its speed. The bunching of the distribution at a multiple of the potential step (Figure 9, right) is the kinetic signature of the regime, and a fluid model cannot draw it. A second case runs the bounded counterpart: a capacitive RF column in argon at 0.1 Torr [10], [11]. There the symmetry forbids a preferred direction, and the four striations stand still, as they do in the laboratory. The ripple of the potential is 5.5 V, not related to any threshold, and the distribution keeps its shape from one striation to the next: a case where a fluid dispersion relation should do better.

### 3.6 The E×B Electron Cyclotron Drift Instability of Hall thrusters

In a Hall thruster, a radial magnetic field holds back the electrons while the axial electric field accelerates the ions. The electrons drift in the azimuthal direction at the velocity E/B, and they cross the magnetic field toward the anode much faster than collisions with the gas can explain. The missing transport is attributed to the electron cyclotron drift instability (ECDI): an ion-acoustic wave that propagates along the drift, and that the magnetised electrons feed at the resonances $\omega - k\, v_E = m\, \omega_{ce}$. A one-dimensional periodic box along the azimuth can contain this instability, on two conditions: the particles must be renewed after a virtual axial transit — the method of Lafleur, Baalrud and Chabert [12] — and the box must hold a whole number of resonant wavelengths $\lambda_1 = 2\pi\, v_E/\omega_{ce}$, the distance an electron drifts during one gyration. The reference run follows the one-dimensional azimuthal simulations of Smolyakov and co-workers [13]: same fields, same density, same gas.

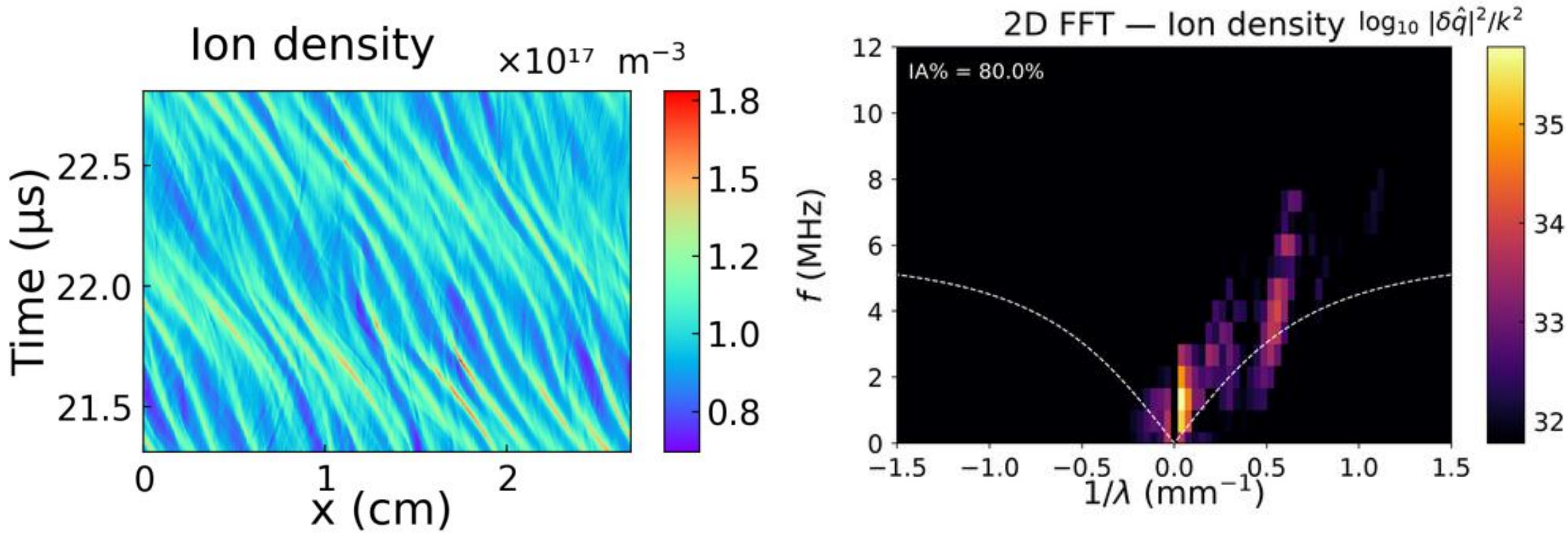

Figure 10. The reference run (xenon, $10^{17}$ m⁻³, B = 200 G, E = 20 kV/m, box of fifteen wavelengths). Left: ion density against azimuthal position and time in the saturated state. The crests move along the electron drift at about 8 km/s, close to the ion-acoustic speed. Right: the two-dimensional spectrum of the ion density; 80 % of the power lies on the ion-acoustic dispersion relation (dashed), and the dominant wavelength, 1.787 mm, is the cyclotron prediction of 1.786 mm.

The series then goes further than a paper usually can. Halving the magnetic field halves the wavelength and the box, and the spectrum peaks again at mode 15. Halving or doubling the density leaves the wavelength exactly where it was, while an ion-acoustic turbulence would have followed the Debye length to modes 10.6 and 21; this is the most direct proof of the cyclotron nature of the instability. Doubling the field, on the other hand, makes the run diverge: the inverse cascade reaches the longest mode of the box within two microseconds, and the electron energy runs away to keV, in the same way with a fixed and with an adaptive time step. The diverged runs are kept in the library for what they are: evidence that the saturated state of a periodic box depends on the box as much as on the plasma, and that thirty wavelengths, not fifteen, are needed at that field. Finally, the ions are seen trapped in the wave crests in phase space, which is what stops the growth.

## 4. Beyond the discharge engine

Two other tools complete the program. A **swarm mode** — a stand-alone Monte Carlo solver of the Boltzmann equation, with no grid, no Poisson equation and no ions, written along the lines of Hagelaar's MCIG — computes the transport and rate coefficients of a gas (drift velocity, longitudinal and transverse diffusion, ionization and attachment, the whole EEPF) as a function of E/N, from the same cross-section files as the discharge engine. The error bars are measured from independent replicas, not modelled. It is the natural tool for checking a set of cross sections, for producing the coefficients a fluid model needs, and for the swarm cases of the library: neon and oxygen sweeps, and a comparison with MCIG.

Two **interactive modules** are intended for a first course. *Single-Particle Motion* integrates the motion of one electron and one ion in prescribed fields and animates the result: gyration; the E×B drift, with both species drifting in the same direction whatever their charge and mass; the grad-B drift; the magnetic mirror and its loss cone; collisions across B with real cross sections; and the oblique-field geometry of the magnetised-sheath cases, on a single particle. *Collisions and Cross Sections* opens the one physical ingredient the code reads rather than computes: the cross sections of a gas, the mean free path and collision frequency they imply, the null-collision method as it is actually executed, the distribution of free-flight times building up under Play, the effect of one collision on a particle, and the rate coefficients. Both modules open from the library on prepared, locked configurations, and can be unlocked for free exploration.

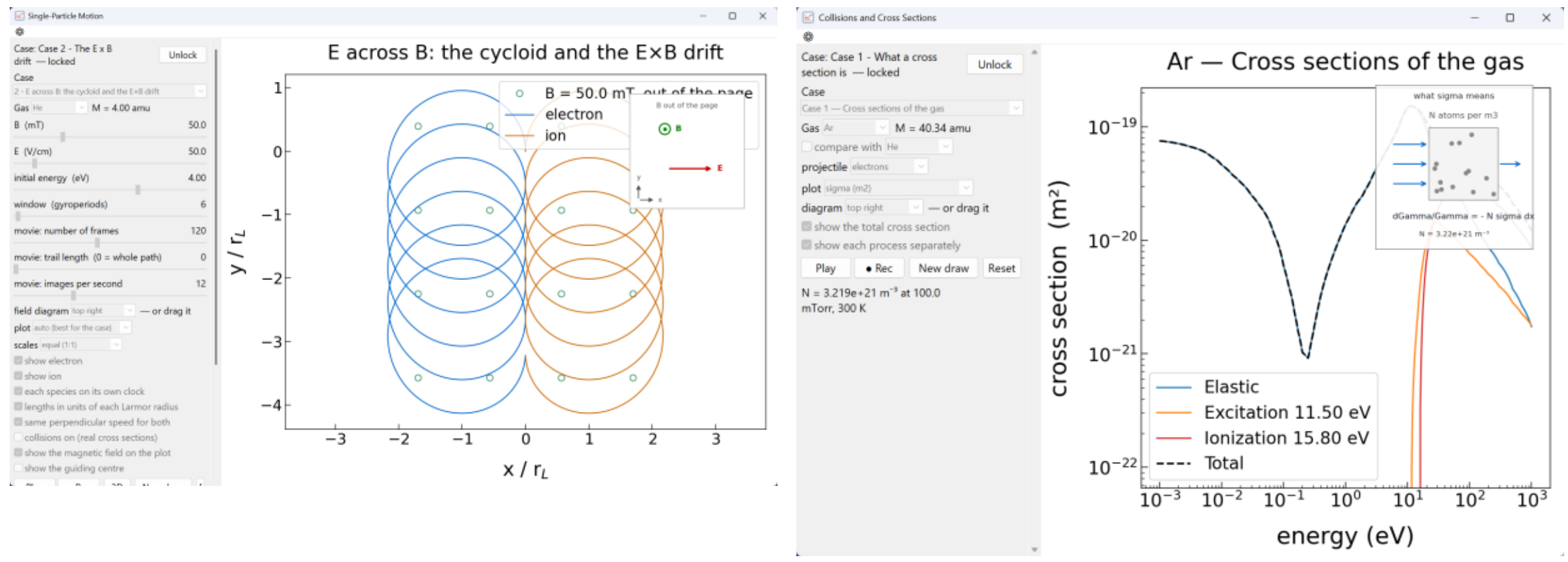

Figure 11. The two interactive modules. Left: Single-Particle Motion on the E×B drift: an electron and an ion gyrate in opposite directions, with radii in the ratio √(M/m), and drift at the same velocity in the same direction. Right: Collisions and Cross Sections on argon, with the Ramsauer minimum near 0.3 eV on which section 3.2 is based.

## 5. Availability, licence and citation

**Availability.** JC-PIC is a Windows program (Windows 10/11, 64-bit), distributed without charge as a single installer from https://jc-pic.org. At the first launch the program asks for a name and an e-mail address and returns an activation key; apart from that step and the optional download of the stored results of the library, it makes no network connection. The user manual is included and works offline.

**Licence.** The software is provided as is, without warranty, under a short end-user licence accepted at installation (no redistribution, no embedding in another product without agreement). The test-case library and the companion book are under CC BY 4.0. The electron cross sections are those of the SIGLO database on LXCat (helium from Biagi-v7.1); the ion cross sections of the rare gases come from the Phelps database.

**How to cite.** Publications using JC-PIC should cite the companion book: J.-P. Boeuf, *Physics of Low Temperature Plasmas via Particle Simulation — the JC-PIC test-case library*, Zenodo (2026), doi:10.5281/zenodo.22258142. The software itself is archived on Zenodo under doi:10.5281/zenodo.22284514, an identifier that always leads to the latest release. It says which program was used; the reference to cite remains the book. The book can also be read online at https://jc-pic.org/library_article.html.

**Contributing.** The library is meant to grow with its users. In the case browser, *Add my case…* turns a working folder into a case of the user's own, with a description to complete. A case worth sharing — a configuration from one's own work, a published result reproduced, a clear illustration of an effect — can be sent to the author. Once integrated, it is published in the program, on the website and in the book, under its author's name.

**Contact.** Questions about the physics, the numerical methods, or a result that does not look right are welcome at jpboeuf@gmail.com. JC-PIC will be presented at the 79th Gaseous Electronics Conference, Chicago, November 2026.

## References

The papers reproduced by the six cases above. The complete bibliography (about 150 entries) is in the companion book.